\documentclass[prl,twocolumn,showpacs,preprintnumbers,amsmath,amssymb]{revtex4}

\usepackage{graphicx}
\usepackage{dcolumn}
\usepackage{bm}

\begin{document}

\preprint{}

\title{Influence of Electron Evaporative Cooling on Ultracold Plasma Expansion}

\author{Truman Wilson, Wei-Ting Chen, and Jacob Roberts}
 
\affiliation{%
Department of Physics, Colorado State University, Fort Collins, CO 80523
}%

\date{\today}

\begin{abstract}
The expansion of ultracold neutral plasmas (UCP) is driven primarily by the thermal pressure of the electron component and is therefore sensitive to the electron temperature.  At lower densities (less than 10$^8$ /cm$^3$), evaporative cooling has a significant influence on the UCP expansion rate.  We studied the effect of electron evaporation in this density range.  Owing to the low density, the effects of three-body recombination were negligible.  We modeled the expansion by taking into account the change in electron temperature owing to evaporation as well as adiabatic expansion and found good agreement with our data.  We also developed a simple model for initial evaporation over a range of ultracold plasma densities, sizes, and electron temperatures to determine over what parameter range electron evaporation is expected to have a significant effect.  We also report on a signal calibration technique, which relates the signal at our detector to the total number of ions and electrons in the ultracold plasma.
\end{abstract}

\maketitle

\section{Introduction}

The creation of ultracold neutral plasmas (UCP) \cite{Killian1999} has opened a new avenue into studies of the physics of strongly-coupled plasmas \cite{Ichimaru1982,Murillo2001,Simien2004,Pohl2004,Chen2004,Denning2009,Niffenegger2011,Bannasch2011,Heilmann2012,Ghosh2012,Lyon2013} through control of the initial energies of the ions and electrons in the system.  Based solely on the initial temperature of the ultracold atoms and the ability to impart little energy during the photoionization process, the ions and electrons in UCPs can be created at relatively low energies, producing plasmas that would have both components in the strongly coupled regime.  However, studies have shown that effects such as disorder induced heating \cite{Denning2009,Heilmann2012,Pohl2005,Gupta2007,Guo2010,Bergeson2011,Lyon2011}, three-body recombination \cite{Gupta2007,Killian2001,Robicheaux2003,Fletcher2007}, threshold lowering \cite{Hahn2001,Hahn2002}, and Debye screening \cite{Chen2004,Lyon2013} can limit the amount of strong-coupling that is realized in UCP systems.  These heating effects increase with the UCP density, though, and this suggests that studying UCPs at low density would likely lead to lower electron temperatures, and in the case of electrons, more strongly-coupled UCPs (the ions would have lower temperatures but the same strong-coupling owing to self-scaling effects).  The effects of electron evaporation also increase with \textit{decreasing} density, further lowering the electron temperature.  This increase in the influence of evaporative cooling with decreasing UCP density is the subject of this work.

The early time electron temperature in UCPs is typically experimentally determined from measurements of the UCP expansion \cite{Gupta2007}, although a direct measurement of this temperature has been reported in \cite{Roberts2004} and the electron temperature has been inferred from three-body recombination rates in \cite{Fletcher2007}.  The finite temperature of electrons will make the electron density slightly lower than the ion density, creating an internal electric field which will drive the ions to expand.  The expansion is typically modelled by a self-similar expansion obtained from the solution to collisionless Vlasov equations, which are applicable through a formal equivalence owing to the spherical symmetry of the UCPs \cite{Laha2007,Morrison2009}.  This expansion follows the form $\sigma (t) = \sqrt{\sigma (0)^2+v^2t^2}$, where $\sigma (0)$ is the initial characteristic size of the UCP from a spherically symmetric Gaussian density distribution given by $n(r) = n_0exp(-r^2/2\sigma (0)^2)$ and $v$ is the asymptotic expansion velocity.  However, this model does not take into account the loss of electrons and the associated energy loss from evaporative cooling.  For many UCP conditions, this energy loss can be negligible, but we find that as the density decreases that this effect becomes significant.  For the UCPs created in our system, the density is 1-2 orders of magnitude lower than what is observed in typical UCP experiments \cite{Killian1999,Simien2004,Bergeson2011,Morrison2009}, which puts our UCPs in a density range where evaporative cooling is expected to be significant.  Thus, evaporation is important for adequate modeling of the UCP temperature evolution as our plasmas expand.  In this manuscript, we develop a model that predicts the effect of electron evaporation over a range of density and temperature conditions.  We also obtained data of UCP expansion, and showed that there is a substantial decrease in the overall expansion energy.  We use our model to show that this reduction in expansion energy is consistent with energy lost via electron evaporation.

The increased importance of evaporative cooling at low density is somewhat counter-intuitive as more robust evaporation is usually associated with higher collision rates and thus higher densities (e.g. ultracold atom experiments \cite{Ketterle1996}).  However for ultracold plasma systems, the self confining nature of the ion/electron system leads to a stronger effect of evaporative cooling for lower density.  A higher density UCP loses a smaller fraction of its electrons to produce the same electron confining potential, limiting the amount of evaporative cooling that occurs.  For instance, if producing a certain amount of confinement requires the loss of 10$\%$ of the electrons in a particular UCP, then for a UCP of the same size but with a factor of 10 increase in the density, only 1$\%$ of the electrons will have to be lost to produce the same amount of confinement.

In order to more quantitatively evaluate the conditions where evaporation may have significant influence on UCP electron temperature evolution and expansion, we constructed a simple model adapted from a cold fluid ion model \cite{Gupta2007,Robicheaux2003,Cummings2005}, modified by adding equations taking evaporation into account \cite{Andresen2010,Twedt2010}.

\begin{eqnarray}
\frac{dN}{dt}=&& -\frac{N}{\tau}\frac{e^{-D/T}}{\sqrt{D/T}}
\\ \nonumber
\\ \frac{dD}{dt}=&& -\alpha\frac{dN}{dt}
\\ \nonumber
\\ \frac{3}{2}Nk_B\frac{dT_e}{dt}=&& (Dk_B-\frac{3}{2}k_BT_e)\frac{dN}{dt}
\\ && -	\frac{3}{4}Nm_{ion}\frac{d}{dt}(\frac{\gamma^2}{\beta}) \nonumber
\\ \nonumber
\\ \frac{d\beta}{dt}=&& -2\beta\gamma
\\ \nonumber
\\ \frac{d\gamma}{dt}=&& \gamma^2 + \frac{2k_BT\beta}{m_{ion}}
\end{eqnarray}
Here $N$ is the number of electrons remaining in the UCP, $D$ is the potential well depth in units of temperature owing to an excess of ions as electrons escape, $\tau$ is the Spitzer electron self-equilibration rate \cite{Spitzer1962}, and $\alpha$ is the change in this potential well depth per electron removed from the UCP.  The $\beta$ and $\gamma$ parameters are those defined in \cite{Cummings2005}, where $\beta = 1/(2\sigma)$ and $\vec{v} =\gamma\vec{r}$, which relates the ion position from the center of the UCP to its velocity.  We integrate equations (1)-(5) until the total energy lost via evaporation becomes approximately constant (changing less than 5$\%$ as the integration time is extended).  In this time period, the size of the UCP increases by less than 10 $\%$.  As the UCP evolves, energy can either be transferred to ion expansion energy or leave through evaporation.  If evaporation does not have a significant effect at early times in the UCP evolution, then as energy goes to ion expansion it is less likely to have a significant effect later in the UCP evolution.  In this model, we do not take into account the effects of three-body recombination.

Equation 1 describes the electron evaporation rate \cite{Andresen2010} from the UCP.  Equation 2 models the change in UCP potential well depth.  $\alpha$ is computed by assuming the starting condition of a spherical UCP whose potential well is just deep enough to trap electrons with kinetic energy equal to the initial ionization energy.  From this starting condition, removing one additional electron increases the potential well depth by $\alpha$, which is dependent on the UCP spatial size and the externally applied electric field.  While $\alpha$ is not truly a constant, because we focus on the time before the size of the UCP increases by 10$\%$, the assumption of constant $\alpha$ is reasonable.  Equation 3 reflects conservation of energy and that the decrease of the electron temperature comes about via both energy transfer to the ions as well as evaporative cooling.  Finally, equations 4 and 5 model the ion expansion.  From the above set of equations, we can calculate the energy loss per ion from evaporative cooling by integrating $dN/dt\cdot (D-3/2T_e)/N$ over the time we are interested in.  Here, the $3/2T_e$ term accounts for the electron loss, allowing us to evaluate the reduction in energy available to drive the ion expansion in the UCP.

Solving these differential equations simultaneously predicts the expansion of the UCP as a function of time under the assumptions that thermal equilibrium is established and maintained and that the expansion is fully self-similar.  This means that the predictions of this model are actually most reliable when the density is high and evaporation is not important as those conditions fulfill these assumptions better.  Thus, equations (1)-(5) are most useful in determining whether or not evaporation is likely to be significant for a given set of UCP conditions.  Determining that evaporation is predicted to be significant is a critical piece of information for interpreting UCP expansion in those cases.

The predictions of this model were explored for UCP parameters at relatively low density, relevant to our experimental parameters, moderate density found in other UCP experiments \cite{Twedt2010}, and high density that was realized by the densities available from the ionization of ultracold atoms in a magneto-optical trap \cite{Gupta2007}.  For each condition the reduction in the amount of energy available for expansion owing to evaporation was evaluated. The results are summarized in Table 1.  As expected, the more quantitative predictions of the model follow the expected relative influence of evaporation on UCP expansion: evaporative cooling is more prominent in the lower density case.  This cooling is a slowly varying function of the density, requiring changes over orders of magnitude in the density to see substantial changes in the cooling owing to the exponential factor in Eq. 1.  These predictions show that influence of evaporative cooling should be substantial for our experimental conditions.  

\small
\begin{table}
\begin{center}
\begin{tabular}{|l|l|l|l|}
\hline
Condition&$\Delta E/k_B$ (K)&Energy&Fraction\\
 & &Removed (K)& \\
\hline
$N_0 = 2.1\cdot 10^5$&25&5.2&0.208\\
$\sigma = 0.8$ mm&50&9.8&0.195\\
$n_{peak} = 2.6\cdot 10^7$ /cm$^3$&75&14.0&0.187\\
\hline
$N_0 = 10^6$&100&10.8&0.108\\
$\sigma = 0.3$ mm&300&33.9&0.113\\
$n_{peak} = 2.4\cdot 10^9$ /cm$^3$& & & \\
\hline
$N_0 = 1.6\cdot 10^8$&21&0.06&0.003\\
$\sigma = 1$ mm&90&1.33&0.015\\
$n_{peak} = 10^{10}$ /cm$^3$&174&0.95&0.005\\
\hline
\end{tabular}
\caption{A summary of the influence of electron evaporation on the available UCP expansion energy over a range of density conditions.  Here we calculate the amount of energy removed from the UCP owing to evaporation that will not be available to drive the ion expansion as described in the text.  The first condition shows the model calculations for experimental conditions achievable in our system.  The second and third conditions show the calculations for the experimental conditions at higher density found in \cite{Twedt2010} and \cite{Gupta2007} respectively.  The model shows that as the density decreases, the influence of electron evaporation on the energy of the UCP increases.}
\end{center}
\end{table}

\normalsize

\section{Apparatus}

In this experiment, the UCPs were created by photoionizing ultracold $^{85}$Rb atoms.  The Rb atoms were first collected and laser-cooled into a magneto-optical trap (MOT) \cite{Chu1998}, then transferred into a magnetic quadrupole trap.  The magnetic trap was created by a set of magnetic coils in an anti-Helmholtz (AH) configuration that were mounted to a motorized translation stage \cite{Lewandowski2003}.  The magnetically trapped atoms were transferred over a distance of $\sim$1 m to the plasma region of the vacuum chamber.  In this region, we have a set of cylindrically symmetric electrodes and a magnetic coil which are described in detail in \cite{Wilson2013}.

After the Rb atoms are transferred, the magnetic trap is turned off and the Rb is ionized in a two-step photoionization process.  The first step is to excite electrons from the $5S_{1/2}$ state in Rb to the $5P_{3/2}$ state using light from a 780 nm diode laser.  To ionize these excited atoms, a pulsed dye laser tuned between 473-479 nm ($\Delta E/k_B = 10-400$ K above threshold) was used.  For the measurements described in this work, the number of total ions was controlled by power of the $5S_{1/2}-5P_{3/2}$ pump laser and the initial electron kinetic energy was controlled by the wavelength of the pulsed dye laser.  After ionization, the electrons are extracted from the UCP to a microchannel plate detector (MCP) using an electric field provided by our electrode assembly.  The electric field provided by our electrodes ranged from $2-10$ V/m.  The electrons striking the MCP ultimately resulted in a current across a load resistance.  This resulted in a voltage that needed to be calibrated in order to determine the number of electrons and ions in the UCP.  A discussion of this calibration as well as the calibration of the electric field is presented in the following section of this Article.
	
The UCPs in this experiment were created with an approximately spherical Gaussian ion and electron distributions described by $n_0exp(-r^2/2\sigma^2)$, where $n_0$ is the peak density and $\sigma$ describes the spatial extent of the UCP.  The UCPs were created with as few as several thousand ions for the calibration to $\sim 5\cdot10^5$ ions for the expansion measurements, with a spatial extent, $\sigma$, that ranged from $700 - 900$ $\mu$m.  This resulted in peak densities that were as high as $\sim 9\cdot10^7$ /cm$^3$ for the UCPs in our experiment.

\section{Number and Electric Field Calibration}

In order to have an accurate measure of the asymptotic expansion velocity to compare to our models, we performed calibration measurements of the MCP signal voltage as it relates to the number of electrons extracted from the UCP.  In principle, the calibration could be performed by comparing the initial number of neutral atoms in the plasma region measured by absorption imaging to our measured MCP signal \cite{Ketterle1999}.  By calculating the ionization fraction based on the intensity of our lasers, a number calibration can be obtained.  However, this method produces relatively large systematic uncertainties.  So instead, we measured the threshold number of ions required to trap electrons in the UCP combined with a measurement of the initial size of the UCP \cite{Killian1999}.  In addition to calibrating the ion number, the calibration technique also allowed the determination of the effective electric field for our electrode voltage configurations used throughout this experiment.
	
The threshold number of ions required to trap electrons in a UCP is defined as the number of ions sufficient to create a potential well with depth, $D$, that is as deep as the initial ionization energy \cite{Killian1999}.  This can be calculated from a Gaussian distribution of ions with a characteristic size, $\sigma$, in an external electric field, $F$.
	
The threshold ion number for a given set of conditions is determined experimentally by measuring the number of electrons trapped in the UCP upon formation as a function of the total number of ions.  We start by forming the UCP under a chosen electric field configuration.  After $\sim$3 $\mu$s, the space charge that develops from the initial escaping electrons forms a potential well trapping any remaining electrons.  At this time, we apply an electric field pulse to extract any electrons trapped in the UCP towards the detector.  We can alter the power of the $5S_{1/2}-5P_{3/2}$ pump laser to control the number of ions in our UCPs without significantly altering the spatial distribution.  As the total number of ions decreases, the number of trapped electrons approaches zero.  We can extrapolate our data to zero trapped electrons for a particular initial ionization energy and electric field ($F$) in order to find the threshold number of ions in units of the MCP signal, $N_{MCP}$.  An example of our MCP calibration data for two different electric fields is shown in Fig. 1.

\begin{figure}
\includegraphics{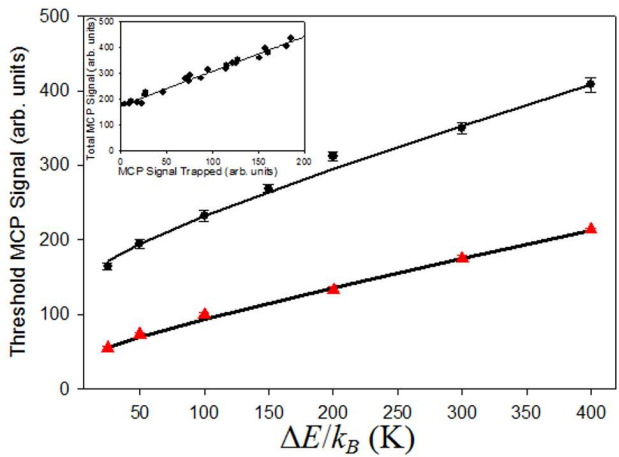}
\caption{(Color Online) The MCP signal calibration measurement uses the MCP threshold signal over range of initial ionization energies for a particular electric field configuration.  The red triangles show the threshold MCP signal for an electric field of 4 V/m, which was found using our calibration.  The black circles show the threshold MCP signal for an electric field of 9 V/m.  The inset figure shows an example of the extrapolation of the trapped electron signal to obtain the MCP threshold signal.  The bottom axis of the inset figure shows the number of trapped electrons and the left axis shows the total number of ions and electrons in units of the MCP signal.}
\end{figure}

To determine the number and electric field calibration, we find the threshold number of ions for several different initial ionization energies for a particular electric field as seen in Fig. 1.  We can multiply the MCP signal by a number calibration factor, $\zeta$, to convert to the actual number of ions in the UCP \cite{Ionnumber}.  We measured the initial peak density of the UCP using the 2-cycle rf sweep technique described in \cite{Wilson2013} in order to link the value of the initial characteristic size of the ion distribution, $\sigma$, to the value of $\zeta$.  For a given value of the electric field, $F$, we can calculate the number of ions required to produce a potential well depth, $D$, for each of the data points in Fig. 1.  We then perform a least squares fit to data to determine the best values of $\zeta$ and $F$.  We performed this calibration technique for several values of the electric field in order to determine the electric field as a function of the electrode voltage and to check the reproducibility of the number calibration.  We were able to determine the value of $\zeta$ with a statistical uncertainty of less than 5$\%$.  Systematic uncertainties associated with measurements of the characteristic size, $\sigma$, and MCP threshold signal are on the few percent level.

\section{Measuring the effect of electron evaporation on UCP evolution}

To measure the UCP expansion, we measure the peak density of the UCP at chosen delay times after the initial photoionization in the UCP evolution using two-cycles of an externally applied rf field as described in \cite{Wilson2013}.  The two-cycles of rf drive a collective oscillation in the electron component of the UCP whose resonant frequency is dependent on the plasma frequency associated with the peak density.  We sweep the frequency of the rf in order to find the maximum response owing to the applied rf, which allows us to determine the peak density of the UCP.

Using our calibration of the MCP signal to the total number of ions in the UCP (taking into account the fraction of electrons that have escaped the UCP \cite{Wilson2013}), we can relate our measurements of the peak density in the center of the UCP to the characteristic size of a Gaussian distribution, $\sigma$, by noting that $\sigma = (N_{ion}/(2\pi)^{3/2}n_{peak})^{1/3}$, where $N_{ion}$ is the total number of ions and electrons in the UCP and $n_{peak}$ is the measured peak density.  By taking these measurements throughout the UCP evolution, we were able to map out the expansion over the full lifetime of the UCP.

If we assume that electron evaporation has no effect on the electron temperature, then all of the initial electron energy should be converted to ion expansion energy as the UCP evolves (Recall that we expect no significant heating from three-body recombination at our UCP temperatures and densities).  For a self-similar expansion, the characteristic size of the UCP would evolve as $\sigma (t)=\sqrt{\sigma_0^2+v^2t^2}$, where $\sigma_0$ is the initial size of the the UCP and $v$ is the asymptotic expansion velocity.  If all of the initial electron energy is converted to UCP expansion energy, then the asymptotic expansion velocity is given by $v = \sqrt{2\Delta E/3m_i}$, where $\Delta E$ is the initial ionization energy and $m_i$ is ion mass.

We compared this self-similar expansion calculation to UCP expansion data taken at $\Delta E/k_B = 25 - 75$ K.  Fig. 2 shows an example of this comparison for over this range of temperature.  For higher initial ionization energies, the assumption of thermal equilibrium is expected to break down for our UCP conditions, as the Spitzer equilibration times \cite{Spitzer1962} can range from several up to 10s of $\mu$s, which can be a significant fraction of the UCP lifetime.  At initial ionization energies much lower than 25 K, the formation of Rydberg atoms will add heat to the UCP.  The dashed lines show the expansion expected from a conversion of all of the initial electron energy to ion expansion energy, which exceeds the observed expansion of the UCP in all cases.  The estimates of our simple model discussed in the introduction indicate that we should expect such a reduction in the expansion rate.  

While our observations confirm the predictions of equations 1-5 that the expansion rate should be lower than what is expected purely from the ionization energy of the electrons, we performed further analysis of the expansion data.  We have several goals for this additional analysis.  First, we aim to confirm that electron evaporation is reasonable explanation for the observed reduction in the UCP expansion rate.  A second goal was to determine whether the reduction in the UCP expansion energy is primarily caused during the formation of the UCP or in the subsequent evolution, which will allow us to more accurately model the UCP expansion.  Finally, we were able to extract the implied electron temperature evolution from this additional analysis.

\begin{figure}
\includegraphics{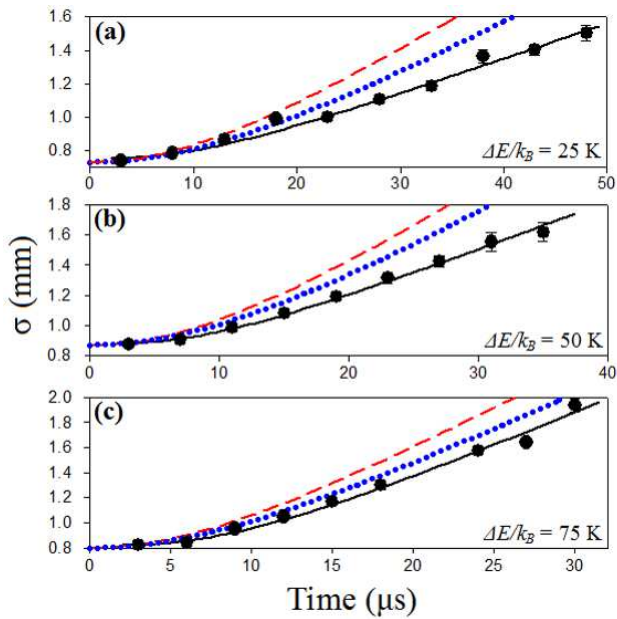}
\caption{(Color Online) An example of the UCP expansion for $\Delta E/k_B = 25-75$ K and $\sim 2.5\cdot 10^5$ total ions.  The data is shown as black circles with error bars representing the statistical uncertainty in our measurement of the UCP characteristic size, $\sigma$.  For clarity the symbol size has been increased, so error bars that are not seen indicate that the uncertainty is less than or equal to the size of the symbol.  The red, dashed line shows a self-similar expansion for all of the energy converted to ion expansion energy.  The blue, dotted line shows a self similar expansion for 20$\%$ of the expansion energy removed during the formation of the UCP.  The solid, black line is a fit of the data taking electron evaporation into account throughout the UCP expansion.  The post-formation average depth to temperature ratio, $D_{avg}/T$ for this calculation was found to be $5.4 \pm 0.4$, $4.7\pm 0.5$, and $2.1\pm 0.2$ for 25, 50, and 75 K to fit our expansion data.  The depths quoted for this data are for electrons assumed to be escaping without any excess energy over the potential well.}
\end{figure}

\begin{figure}
\includegraphics{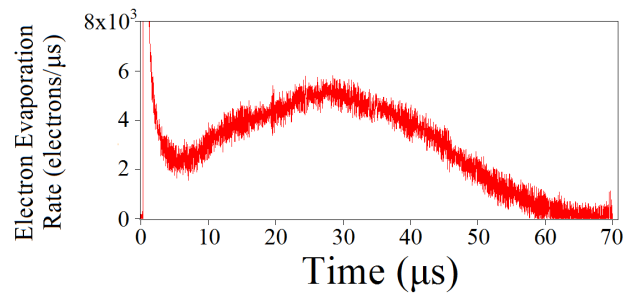}
\caption{An example of our electron evaporation signal for $\Delta E/k_B = 50$ K and $2.5 \cdot10^5$ total ions and electrons.}
\end{figure}

We separate our model of the electron evaporation into two parts, with the first part focusing on the UCP formation and the other focusing on the subsequent expansion.  To improve the accuracy of the model, we use the measured evaporation rate (Fig. 3) in place of Eq. 1 for both parts since the evaporation rate is obtained as part of our data collection.  During formation ($< 2$ $\mu$s after ionization), there will not be significant ion expansion, allowing us to ignore ion dynamics and explicitly calculate the potential depth of the UCP as electrons escape.  As the UCP evolves after initial formation, the outer ion dynamics become complicated.  During this time, we assume a fixed potential depth to electron temperature ratio as the UCP expands to elucidate the general behavior.  Further modeling of the post-formation depth variation including the dynamics of the outer ions is beyond the scope of this work.

The first part of our model calculates the energy removed from the UCP by individual electrons as they escape during the formation stage, which for our UCPs is the first 2 $\mu$s.  Immediately after photoionization, the kinetic energy of each electron will be approximately equal to the initial ionization energy.  In the presence of an extraction electric field, a minimum number of electrons are required to escape before the resulting charge imbalance begins to create any theoretical confinement, even for zero-kinetic-energy electrons.  As a charge imbalance develops owing to escaping electrons, a potential well will form that begins to confine the remaining electrons in the UCP.  At this point, some degree of electron thermalization must take place, as the potential well barrier can become greater than the initial ionization energy.  In our model of evaporation during UCP formation, we assume that as a potential well begins to form, the electrons that escape will have a fixed energy above the total depth of the potential well.  As the electrons escape, they carry away more energy than the average thermal energy of the remaining electrons, lowering the overall temperature.

To calculate the potential well depth, $D$, we used a $T=0$ electron and ion distribution in the presence of a uniform electric field.  The ion distribution was assumed to be a spherically symmetric Gaussian distribution given by $n_0exp(-r^2/2\sigma^2)$, where $n_0$ is the peak density and $\sigma$ describes the characteristic size of the UCP.  We calculated the potential depth of the UCP by integrating from the center of the plasma the electric field produced by the UCP and the external field, under the assumption that the external field is completely screened by the UCP electrons.  To account for the offset of the electron cloud of the UCP from the ions owing to the external electric field, a dipole electric field term was added which maintained the internal screening of the UCP.

\begin{figure}
\includegraphics{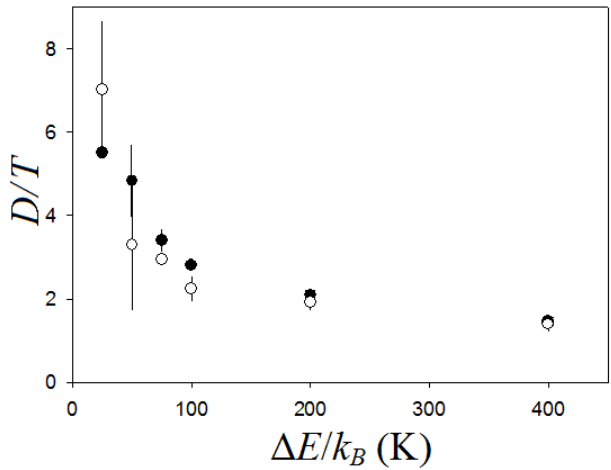}
\caption{(Color Online) A plot of the calculated depth to temperature ratio, $D/T$ over a range of temperatures.  The open/closed circles show data for a total ion number of $\sim 2.5/5.0\cdot 10^5$ ions respectively.  The vertical lines for this data represent the range of depths that were calculated for $k_BD$ to $k_BD+\Delta E$ of energy removed by each escaping electron for multiple sets of data as described in the text.  The depth to temperature ratio shows a sharp decrease at low temperatures that levels off at higher temperatures.}
\end{figure}

The electron evaporative cooling rate as a function of time was taken from our experimental data over a range of initial ionization energies ($\Delta E/k_B = 25 - 400$ K), total number of ions (($2.5-5.4)\cdot 10^5$) and $\sigma \sim$ 750 $\mu$m.  This part of the model makes no assumptions about thermal equilibrium, and is therefore valid over a broader range of initial ionization energy.  We calculated the effect of the the electron evaporation in the UCP prompt peak by numerically evaluating how much energy was removed during each time step of the UCP formation process, using our measured evaporation rate (Fig. 3).  For each time step, we calculated the new potential depth, $D$, based on the total amount of charge that has left the UCP up to that point.  For $D<0$ (i.e. no potential well formed), the escaping electrons are assumed to leave the UCP with their initial energy from ionization.  This part of the formation typically lasts $<$100 ns, which is less than the electron-electron thermalization times \cite{Spitzer1962} for all of our UCP conditions.  For $D>0$, the escaping electrons must have kinetic energies $\geq k_BD$.  For this portion of the formation process, we calculated the amount of energy that escaped the UCP for electrons with energies that ranged from $k_BD$ to $k_BD+\Delta E$, representing a reasonable range of likely electron escape energies.

We calculated the amount of total energy that left the UCP through evaporation, divided the remaining energy by the number of electrons remaining in the UCP and compared this to the initial ionization energy.  The results of these calculations showed that for the lower end of our total ion number range, the amount of energy available for expansion is reduced by $15-25\%$ over our range of temperatures.  At higher total ion number, this energy is reduced by $8-12\%$.  As the initial ionization energy decreases, the fractional amount of energy removed during the formation process increases.  To show this, we calculated the depth to temperature ratio, $D/T$, which showed that at low temperatures this ratio increases (Fig. 4), which means the energy removed per electron is higher when compared to the average energy of the electrons in the UCP.  The decrease in $D/T$ with increasing ionization energy is consistent with the expected decrease in thermalization rates for the UCPs in our system.

In some experimental conditions, electron evaporation after the UCP formation can also have a significant effect on the UCP expansion.  Contrary to the case of the UCP formation, the expansion dynamics of the outer ions become important and complicate any attempt to model the potential well depth of the UCP with precision.  In the absence of a sophisticated model of the ion dynamics, we can still observe the relative importance of post-formation evaporative cooling by assuming a constant potential well depth to temperature ratio, $D/T$.  In reality, this ratio will vary somewhat in time as the UCP expands.  In this treatment, the assumption of constant $D/T$ is equivalent to a weighted average value over the UCP expansion.  The value of $D/T$ for any given set of experimental conditions is determined through fitting the expansion data to find the best-fit value.

\begin{figure}
\includegraphics{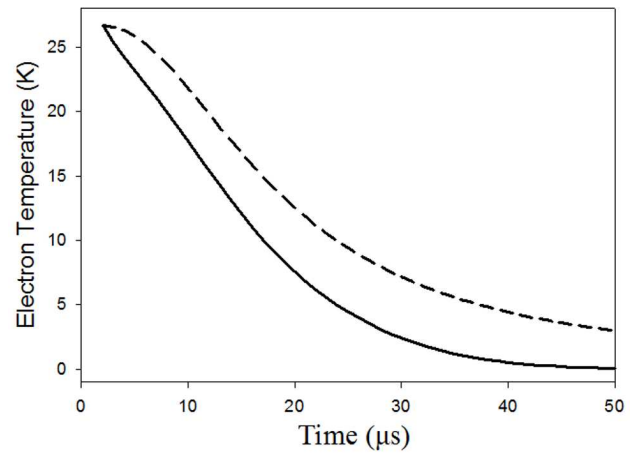}
\caption{The electron temperature evolution of an expanding ultracold plasma with (solid) and without (dashed) evaporation for $\Delta E/k_B = 50$ K.  Here we are looking at the temperature evolution after the UCP formation (starting at 2 $\mu$s, where 20$\%$ of the energy has been removed).  The solid line shows the electron temperature evolution corresponding to the solid, black line from Fig. 2, whereas the dashed curve shows the temperature evolution for the blue, dotted line in Fig. 2.  It is clear that evaporation lowers the temperature of the electron component at a much faster rate at early times in the evolution, which affects the expansion at later times.}
\end{figure}

After performing this calculation for different initial ionization energies, the averaged depth to temperature ratio, $D/T$, was $5.4 \pm 0.4$, $4.7 \pm 0.5$, and $2.1 \pm 0.2$ for $\Delta E/k_B$ = 25, 50, and 75 K respectively as seen in Fig. 2.  This trend is consistent with the $D/T$ deduced from UCP formation calculation.  The values of $D/T$ calculated from our expansion data are in a range consistent with what is expected from our observed evaporation rate, indicating that evaporation is a reasonable explanation for the observed expansion rate reduction.  From this calculation, we were also able to infer the temperature evolution of the UCP under these conditions as seen in Fig. 5.  We can see from this figure that the reduction in the electron temperature as a function of time is significantly different from the case where no evaporation is present.  We also found that as the number of total ions and electrons is increased from $2.5\cdot 10^5$ to $5\cdot 10^5$ that the expansion of the UCP does not change significantly.  This is consistent with the expectation of our model that the effect of evaporation varies slowly with the density.

The temperature evolution of the UCP deduced from our model (Fig. 5) combined with our measurements of the expansion also allow us to calculate the strong-coupling parameter, $\Gamma$ \cite{Ichimaru1982}, over the lifetime of the UCP in the absence of three-body recombination.  These calculations show that the strong-coupling parameter could become greater than the theoretical limit of 0.2 \cite{Robicheaux2002}.  This would occur at a time in the evolution when the expected heating owing to three-body recombination is still small.  While suggestive, the fact that small heating rates would not easily be observed in our data at late times in the UCP evolution means that a definitive determination of $\Gamma$ requires adding additional capabilities to our system.  This includes the ability to measure the Rydberg atom formation rate, and to have a local electron temperature probe to more accurately measure the temperature \cite{Roberts2004}.  These additional experimental capabilities would also facilitate an investigation of the effectiveness of forced evaporative cooling to lower the electron temperature at earlier times.  These studies will be the focus of future work.

\section{Conclusion}

In summary, we have observed a substantial decrease in UCP expansion rate owing to electron evaporation on the expansion of ultracold plasmas in our system.  We have performed calculations that show evaporation is expected to have a greater effect for low UCP density conditions, consistent with results obtained from our system.  The observed expansion rate was significantly lower than that of a self-similar expansion obtained from a solution to the collisionless Vlasov equations with the full initial ionization energy.  To further characterize this loss of expansion energy owing to evaporation, we performed calculations using electron evaporation data over a range of conditions to determine the amount of energy lost in the formation of the UCP and in the UCP expansion.  These calculations showed that expansion energy is lost both during formation and from electron evaporation as the UCP expands.  Our data shows that for our experimental conditions, that evaporative cooling has a significant effect on electron temperature reduction.  This leads to the possibility of forced evaporative cooling to temporarily raise the value of the strong coupling parameter of the electron component of UCPs, which will be the subject of future work.

This work was supported by the Air Force Office of Scientific Research, Grant No. FA 9550-12-1-0222.


\begin{thebibliography}{1}
\bibitem{Ionnumber} There is one free electron produced for every ion that is created, so measuring the total electron number gives the total ion number.
\end{thebibliography}
\end{document}